\documentstyle[12pt]{article}
\begin{document}

\begin{center}
{\LARGE{\bf COHERENT PION PRODUCTION \\
\vspace{0.3cm}
IN NEUTRINO NUCLEUS COLLISION IN THE 1 GEV\\
\vspace{0.3cm}
  REGION}}
\end{center}

\vspace{1cm}

\begin{center}
{\large{N. G. Kelkar$^{(1,2)}$, E. Oset$^{(1)}$ and
 P. Fern\'andez de C\'ordoba$^{(3)}$}}
\end{center}

\vspace{1cm}

{\small{\it
\noindent
$^1$ Departamento de F\'{\i}sica Te\'orica and IFIC, Centro Mixto 
Universidad de Valencia-CSIC, 46100 Burjassot (Valencia), Spain.\\
\noindent
$^2$ Laboratori Nazionali di Frascati, INFN, Italy\\
\noindent
$^3$ Departamento de Matem\'atica Aplicada, Universidad Polit\'ecnica
de Valencia}}

\vspace{2cm}

\noindent
PACs numbers: 25.30-c, 25.30. pt

\vspace{3cm}

\begin{center}
\begin{abstract}
{\small{We calculate cross sections 
for coherent pion production in nuclei induced by
neutrinos and antineutrinos of the electron and muon type. The 
analogies and differences between this process and the related ones 
of coherent pion production induced by photons, or the $(p,n)$ and 
$(^3 He, t)$ reactions are discussed. The process is one of the 
several ones occurring for intermediate energy neutrinos, to be 
considered when detecting atmospheric neutrinos. For this purpose 
the results shown here can be easily extrapolated to other energies 
and other nuclei.}}
\end{abstract}
\end{center}

\newpage
\section{Introduction}
Electron and muon neutrinos and antineutrinos at intermediate 
energies are produced in cosmic ray interactions with the earth's 
atmosphere and they are classified as ``atmospheric neutrinos''. From 
basic counting from pion and muon decays one expects twice as 
many muon neutrinos and antineutrinos as the corresponding electron 
ones, which seems to be in contradiction with measurements at IMB 
\cite{BEC} and Kamiokande \cite{HIR}, where the ratio obtained by 
studying charged current neutrino nucleus reactions in large 
underground water detectors is about one. One of the attractive 
hypothesis is the existence of neutrino oscillations. However, firm 
conclusions on the reasons of the puzzle can only come if we have a 
good control on the neutrino nucleus reactions occurring in the 
detectors, as well as the detector characteristics.

The neutrino nuclear reactions at intermediate energies can be rather
complicated if one compares with analogous reactions induced by 
photons \cite{CAR}. Most of the studies of neutrino nucleus 
collisions only consider the ph excitation channel
\cite{HAX,KUB,ENG}. Others include $\Delta h$ excitation as a source 
of renormalization of the $ph$ excitation channel but not as an 
excitation channel by itself \cite{SIN,KOS}. All these approaches 
are fine in order to evaluate the nucleon emission channel, and at 
energies below 300-400 MeV they can also provide accurate results for 
the total neutrino cross section. Indeed, most of the neutrino energy 
is transferred to the lepton (in charged currents reactions) and only 
a fraction of it is used to excite the nucleus. Hence, one has to go 
to relatively large neutrino energies in order to excite $\Delta$'s 
and other resonances in the nucleus. Yet, whenever this happens the 
cross sections for $\Delta h$ excitation are large and comparable in 
size with those of ph excitation. A recent evaluation of these 
cross sections is done in ref. \cite{KIM} and the astrophysical 
consequences of the consideration of these channels are discussed in 
ref. \cite{HKI}.

One of the interesting channels discussed in ref. \cite{HKI} is the
coherent pion production following $\Delta$ excitation in the 
nucleus. The reaction is

\begin{equation}
\nu_l  + A \rightarrow l^- + A + \pi^+
\end{equation}

\noindent
where the final nucleus is the same as the original one and it is 
left in its ground state. The cross section for the reaction is 
evaluated in \cite{HKI} using the impulse approximation (IA), 
neglecting the $\Delta$ renormalization in the medium and using 
plane waves for the pion. The authors, however, make a call for 
accurate calculations which would take those elements into account. 
Fortunately, such elements have been thoroughly
tested in pion nuclear reactions and are readily available. The
$\Delta$ selfenergy in a nuclear medium has been evaluated
theoretically \cite{OSE} and tested in all sorts of pion nuclear 
reactions: elastic \cite{GAR} absorption, inclusive quasielastic,
charge exchange and double charge exchange \cite{SAL}, as well as in
pion production processes like coherent $\pi^0$ photoproduction 
\cite{RCA} or coherent pion production in $(p, p')$ and related 
reactions \cite{FER}.

Coherent pion production in $(p,n)$ or $(^3 He,t)$ reactions 
ressembles much the present process and the findings of those 
reactions can serve as a guideline for the study of the present 
one. Coherent pion production in the $(^3 He,t)$ reaction has 
received most of the attention \cite{OLT,PFE,DIM} since it is one of 
the channels contributing to the inclusive  $(^3 He,t)$ reaction in 
nuclei, where an apparent shift of the $\Delta$ peak with respect to 
the peak of the elementary reaction on proton targets was observed 
\cite{ELL,ABL}. In fact the coherent pion production channel shows 
the peak of the $\Delta$ shifted towards lower excitation energies 
in all the calculations \cite{OLT,PFE,DIM},
something already observed in pion elastic scattering as a 
consequence of the pion multiple scattering \cite{ERI}.

Coherent pion production in neutrino reactions can offer additional 
information over the $(p, p'), (^3 He, t)$ and $(p,n)$ reactions on 
$\Delta$ properties and pion propagation
in the medium. The reason is
that in neutrino reactions the range of energy and momentum 
transferred to the nucleus is different than the one in the hadron 
induced reactions. On the other hand in the hadronic reactions one 
has to fight the distortion of the nucleons, or $^3 He, t$ in their 
passage through the nucleus, which makes the pion production reaction 
quite peripheral. Instead, in the neutrino induced reaction neither 
the neutrino nor the lepton is distorted by the nucleus and one can 
test the $\Delta$ and the pions in the interior of the nucleus. One 
may argue that the same occurs in coherent $\pi^0$ photoproduction.
However, in this latter case the combination of the spin transverse
photons and the spin longitudinal pions leads to a factor 
$\sin \theta$ in the amplitude which eliminates the contribution of 
small angles to the cross section. As a consequence the cross 
section picks up its strength from finite angles where the momentum 
transfer is larger and hence the nuclear form factor smaller. This 
reduction due to the form factor becomes more apparent as the energy 
increases, for a given angle, and as a result of this the $\Delta$ 
peak is shifted to much smaller energies than in coherent pion 
production induced by $(p,p'), (^3 He, t)$, etc. The contribution 
of bigger densities is partly to be blamed for the shift, but to a 
much  smaller extent than the reasons discussed above.

In the neutrino induced coherent pion production reaction we do not 
have the circumstances explained in the photon case and, as we shall 
see, the largest contribution to the cross section comes from small 
angles. Then the situation is rather different and we can obtain new 
information with respect to both the photonuclear and strong 
interaction induced processes of coherent pion production.

In this sense the neutrino induced coherent pion production
reaction is a nice complement to other existing reactions which can
 enrich our understanding of the nuclear excitation mechanisms at
intermedite energies providing new tests for the present theoretical
models.

We shall follow the steps and the formalism developed in refs. 
\cite{FER,PFE}. Although accurate experimental
data on coherent pion production from the $(p, p')$ or
$(^3 He, t)$ reactions do not exist, the results of ref. \cite{PFE} 
and the preliminary data of ref. \cite{HEN} for the $(^3 He, t)$
induced reaction are in relatively good agreement, which gives us a 
certain confidence to extend the method to the neutrino induced
reaction.

\section{Coherent pion production amplitude}

The experiments on $(^3 He,t)$ induced pion production 
\cite{ELL,ABL,HEN} show that the process is dominated by $\Delta h$ 
excitation, even at energies
of the beam of 10 GeV. The excitation of other resonances is
suppressed with respect to the $\Delta$.
Only in experiments where the $\Delta$ excitation on the target is
forbidden, as in the $(\alpha, \alpha')$ reaction on a proton target,
has the Roper some chances to show up \cite{MAR}, and even then, the 
Roper signal is small compared to a large background of $\Delta$ 
excitation in the projectile \cite{MAR,PFC,SHI}. The coherent 
production process still restricts more the excitation of higher 
resonances since this requires larger momentum transfers which make 
the nuclear form factor smaller.

The mechanism for coherent pion production in the $(\nu, l^-)$ 
induced reaction proceeds as shown in fig. 1. The $W^+$ emitted from 
the $(\nu,l^-)$ vertex excites a $\Delta$ in the nucleus, which 
decays to $N \pi$ later  on, the nucleon remaining in the same 
original state in order to ensure the coherence.

In order to construct the amplitude for the process we recall the 
$\nu n \rightarrow l p$ weak interaction Lagrangian

\begin{equation}
L = \frac{G}{\sqrt{2}} \cos \theta_c l^\mu J_\mu
\end{equation}

\noindent
with

\begin{equation}
\begin{array}{ll}
l^\mu =  & \bar{u} (k') \gamma^\mu (1 - \gamma_5) u (k)\\[2ex]
J_\mu = & \bar{u} (p') [F_1^{(v)} (q^2) \gamma_\mu + \frac{i}{2M}
F_2^{(v)} (q^2)  \sigma_{\mu \nu} q^\nu \\[2ex]
& + F_A^{(v)} (q^2) \gamma_\mu \gamma_5 + F_P^{(v)} (q^2) q_\mu 
\gamma_5] u (p)
\end{array}
\end{equation}

\noindent
where $G$ is the Fermi weak coupling constant, $M$ is
the nucleon mass and $\theta_c$ is the Cabbibo angle.
We follow the nomenclature and use the same form factors as in 
ref. \cite{SIN}. The momenta involved in eqs. (3) are depicted in
fig. 2.

In order to construct the $\nu N \rightarrow l^- \Delta$ transition 
we make the nonrelativistic reduction of the terms in eqs. (3), 
neglecting only terms of order $ O (\frac{p}{2M})^2$.
 Linear terms in $\frac{\vec{p}}{2M}, \frac{
\vec{p}\,'}{2M}, \frac{\vec{q}}{2M}$ are kept, but those terms 
linear in $\vec{p}$, the momentum of the occupied nucleons, also 
give rise to $O (\frac{p}{2M})^2$ corrections when integrating over 
the momenta. Hence these terms are dropped, which is equivalent to 
taking $\vec{p}= 0, \vec{p}\,' = \vec{q}$. For the 
$\nu N \rightarrow l^- \Delta$
transition we take the terms involving spin operators in the 
$\nu N \rightarrow l^- N$ transition and make the 
substitution

\begin{equation}
\frac{f}{\mu} \sigma^i \tau^\lambda \rightarrow
\frac{f^*}{\mu} S^{\dagger \, i} T^{\dagger \, \lambda}
\end{equation}

\noindent
where $S^i, T^\lambda$ are the spin, isospin transition operators
from 1/2 to 3/2, normalized as

\begin{equation}
< 3/2 \; \, 
M_s | S_\mu^\dagger | 1/2 \; m_s > = C (1/2 \; \;  1 \; \, 3/2 \, ;  
m_s , \mu , 
M_s)
\end{equation}

\noindent
and the same for $T^{\dagger \, \lambda}$. The couplings $f, f^*$
correspond to the $NN \pi$ and $N \Delta \pi$ vertices,
$f^2 /4 \pi = 0.08 , \; f^{*2}/4 \pi = 0.36$ and
$\mu$ in eq. (4) is the pion mass. The current $J_\mu$ of eq. (3)
contains implicitly a factor $\sqrt{2}$ of the operator $\tau_+$
responsible for the $n  \rightarrow p$ transition.

Hence we must substitute

\begin{equation}
\sigma^i \rightarrow \frac{f^*}{f} \frac{1}{\sqrt{2}} 
S^{\dagger \, i} \, T^\dagger_{+}
\end{equation}

\noindent
and from the $\Delta $ decay into $\pi N$ we take $f^* /f = 2.12$, 
a value in between the factor 2.2 taken in ref. \cite{DEK} and the 
factor 2 considered in ref. \cite{KIM}

We find

\begin{equation}
\begin{array}{l}
J_\mu^\Delta = \frac{1}{\sqrt{2}} \frac{f^*}{f} \left\{  \right.
i[ F_1^{(v)} (q^2) + F_2^{(v)} (q^2) ] \frac{1}{2M} (\vec{S}\,
^\dagger 
\times \vec{q})^i \delta_{\mu i}\\[2ex]
+ [ F^{(v)}_A (q)^2 - q^0 F_P^{(v)} (q^2) ] \frac{\vec{S}\,^\dagger
\vec{q}}{2M} \delta_{\mu o} - F_p^{(v)} (q^2)  \frac{\vec{S}\,
^\dagger
\vec{q}}{2M} q^i \delta_{\mu i}\\[2ex]
+ F_A^{(v)} S^{\dagger i} \delta_{\mu i} \left. \right\} 
T^\dagger_{+}
\end{array}
\end{equation}

For the $\Delta $ coupling to a nucleon and a real pion in fig. 1 we 
use the standard vertex

\begin{equation}
- i \delta \tilde{H} = - \frac{f^*}{\mu} \vec{S} \cdot \vec{p}_\pi 
T^{\lambda}
\end{equation}

\noindent
where $\vec{p}_\pi$ is assumed in the $\Delta$ CM frame. The 
amplitude corresponding to fig. 1 is readily evaluated for a 
spin saturated nucleus using the property

\begin{equation}
\sum_{M_s} S_i S_j^{\dagger} = \frac{2}{3} \delta_{ij} -
\frac{i}{3} \epsilon_{ijl} \sigma_l
\end{equation}

One of the findings in coherent pion production induced by the
$(^3He, t)$ reaction was the negligible contribution from the
transverse part of the $NN \rightarrow N \Delta$ transition 
amplitude. This occurred because the emitted pion couples 
longitudinally to the $\Delta$ and the transverse part of the 
interaction contributes to the cross section with the factor 
$\sin^2 \theta$, as we indicated for the case of coherent pion 
photoproduction. This forces the contribution at finite angles 
where the nuclear form factor reduces the cross section. This is 
also the case here, where in addition the transverse terms are 
further reduced  by a
factor $q/2M$. Hence we neglect the transverse parts from the
beginning.

With all these considerations, the coherent pion production cross
section corresponding to the process of fig. 1, on summing over all
occupied nucleons in the amplitude, is given by 

\begin{equation}
\frac{d \sigma}{d \Omega_e d E_e d \Omega_\pi} = \frac{1}{8}
\frac{|\vec{k}'| |\vec{p}_\pi|} {|\vec{k}|} \frac{1}{(2 \pi)^5} 
\Pi_f 2 m_f \bar{\Sigma} \Sigma | t |^2
\end{equation}

\noindent
where the product of the fermion masses, $2 m_f$, appears because
of our normalization of the spinors as $\bar{u} u = 1$.
 The $T$ matrix 
squared,  summed and averaged over spins, is given by

\begin{equation}
\Pi_f 2 m_f \bar{\Sigma} \Sigma |t|^2 = L_{00} | V^0|^2 +
L_{33} |V^3|^2 + 2 L_{03} Re \{ V^0 V^{3*} \}
\end{equation}

\noindent
where $L_{\mu \nu}$ is the leptonic tensor

\begin{equation}
L_{\mu \nu} = 4 [k_\mu k'_\nu + k'_\mu k_\nu - k \cdot k' 
g_{\mu \nu}]
\end{equation}

\noindent
and

$$
V^\mu \equiv \left\{
\begin{array}{c}
V^0\\
0\\
0\\
V^3 \end{array} \right\}
$$

\noindent
\begin{equation}
\begin{array}{l}
V^0 = B [F_A^{(v)} (q^2) - q^0 F^{(v)}_P (q^2) ] \\[2ex]
V^3 = B [F_A^{(v)} (q^2) \frac{2M q}{\vec{q}\,^2}
 - F^{(v)}_P (q^2) q ] \\
\end{array}
\end{equation}

\noindent
with

\begin{equation}
B = - \frac{f^*}{f} \frac{f^*}{\mu}  \frac{G}{6}
\cos \theta_c \, G_\Delta (p_\Delta) F (\vec{q} - \vec{p}_\pi)
\frac{1}{\sqrt{s_\Delta}}
\end{equation}

In the factor $B$, $G_\Delta (p_\Delta)$ is the $\Delta$ propagator
and $F (\vec{q} - \vec{p}_\pi)$ is the nuclear form factor 
modulated by the isospin factors. We have

\begin{equation}
G_\Delta (p_\Delta) = \frac{1}{\sqrt{s_\Delta} - M_\Delta + i
\frac{\tilde{\Gamma}}{2} - \Sigma_\Delta }
\end{equation}

\noindent
where $s_\Delta = p^{0 2}_\Delta - \vec{p}_\Delta^{\,2}$, and 
$\tilde{\Gamma}$, $\Sigma_\Delta$ 
are the Pauli blocked $\Delta$ width and the rest of the $\Delta$
selfenergy which contains pieces
related to quasielastic scattering, $2N$ and $3N$ pion absorption. 
The evaluations are done in ref. \cite{OSE} and we take the
analytic expressions derived there. We also include 
in the selfenergy
the term $\frac{4}{9} 
(\frac{f^*}{\mu})^2 g' \rho$ to account for irreducible pieces of
$\Delta h$ propagation mediated by the Landau-Migdal effective 
interaction \cite{GAR}.

On the other hand the nuclear form factor is given by

\begin{equation}
F (\vec{q} - \vec{p}_\pi) = \int d^3 r [ \rho_p (\vec{r}) + 
\frac{1}{3} \rho_n (\vec{r})] e^{i \vec{q} \cdot \vec{r}} 
\vec{p}_\pi \cdot \vec{q} 
e^{- i \vec{p}_\pi \cdot \vec{r}}
\end{equation}

\noindent
where for convenience we have included the factor 
$\vec{p}_\pi \cdot \vec{q}$.

In the derivation of eq. (11) we have taken $\vec{q}$ in the z 
direction for simplicity and furthermore we have also kept only 
the longitudinal part of $\vec{p}_\pi$ along the $q$ axis for 
consistency with the neglect of the transverse parts. The structure 
of eq. (11) is also the same as the one found in ref. \cite{HKI}.

It is interesting to note that when $\vec{k}'$ is paralel to 
$\vec{k}$, which 
leads to  the largest cross sections, the contribution of 
$F_P$ cancels. Hence,
 the axial term $F_A^{(v)} (q^2)$ is the relevant term
in the process.

So far the formalism has used the bound wave functions of the 
nucleus, which appear in the nuclear form factor via the proton 
and neutron densities, eq. (16), but has considered only a plane 
wave for the pion. The  renormalization of the pion is a very 
important thing in this process. Hence, in the next step we replace

\begin{equation}
\vec{p}_\pi \cdot \vec{q} \; e^{-i \vec{p}_\pi \cdot \vec{r}} 
\rightarrow
i \vec{q} \cdot \vec{\bigtriangledown} \phi_{out}^* (\vec{p}_\pi, 
\vec{r}) 
\end{equation}

\noindent
where $\phi^{*}_{out} (\vec{p}_\pi, \vec{r})$ is an outgoing 
solution of the Klein Gordon equation for the pion, which we solve 
along the lines of ref. \cite{NIE} and with the pion nucleus optical
potential developed there, which gives rise to good pion elastic,
absorption and quasielastic cross sections \cite{NIE}.

\section{Results and discussion}
In fig. 3 we plot $d \sigma / d \Omega_e d E_e d \Omega_\pi$, for 
$\theta_e = 0^0$ with respect to the neutrino direction, for coherent
pion production in 
$\nu_e +\, ^{16}O \rightarrow e^- +\, ^{16}O + \pi^+$ as a function 
of the pion angle measured with respect to $\vec{q}$. We
choose a neutrino energy of $800 \, \,MeV$ and an electron energy of 
$545 \, \,MeV$
which lead to a value of $q$ suited to excite the $\Delta$ resonance. 
The dashed line corresponds to the impulse approximation, meaning 
free $\Delta$ width and no $\Delta$ selfenergy in the $\Delta$ 
propagator and no pion distortion. The solid line is the accurate 
calculation, which accounts for both effects. We can see that there 
is a net reduction of about a factor three from both 
renormalizations, bigger than what appears in coherent pion 
production induced by the $(^3{He}, t)$ reaction, which is more 
peripheral.

The cross section is forward peaked, as was also the
case in the $(^3{He}, t)$
or $(p, n)$ reactions. However the fall down with angle is not so 
drastic here as in the hadronic reactions because in the latter ones, 
for the same energy of the pion, the momentum transferred to the 
nucleus is bigger than in the neutrino case, as a consequence of the 
large mass of the nucleons, and the nuclear form factor reduces more 
the cross section. The cross sections are of the order of 
$10^{-15} fm^2 / MeV sr^2$.

In fig. 4 we show the cross section for the same reaction integrated 
over the pion angles. Here we plot it as a function of $q^0$, the 
total pion energy (we neglect the nucleus recoil energy). We observe 
again the sizeable renormalization factor from dressing the $\Delta$ 
and the pion in the nuclear medium. It is also worth looking at the 
shift to lower energies of the peak of the excitation function, with 
respect to the one of the impulse approximation. This is mostly due 
to the distortion of the pion, as we already indicated 
in the introduction. The argument
goes as follows: since the $\pi N$ cross section and pion absorption 
are largest at resonance, there is a depletion of the pion wave 
when the pion goes through the nucleus, and much of the pion flux is 
lost into quasielastic channels or pion absorption. On the other hand 
the pion production step is resonance peaked. The combination of 
these two factors has as a consequence a shift of the peak to lower 
excitation energies where the pion depletion is not so strong.
Note, however, that the free $\Delta$ position
would appear at $q^0 = 338 \; \,MeV$ in the plot. Hence, we see a 
shift of the peak already in the IA due to the nuclear form factor, 
as indicated in the introduction, and a further shift due to the pion 
distortion.

The nuclear form factor acts as follows: from energy conservation we 
have $\omega_\pi = q^0$, $p_\pi = (q^{02} - \mu^2)^{1/2}$. Hence, the 
momentum transfer to the nucleus, $\vec{q} - \vec{p}_\pi$, is always 
finite since $q > q^0$ and it increases as $q^0$ increases. As a 
consequence, the nuclear form factor decreases with increasing $q^0$ 
and this  has the same effect as the distortion when one approaches 
the $\Delta$ energy, leading to a shift of the $\Delta$ peak to lower 
excitation energies.

The $\Delta$ peak in fig. 4 appears around $q^0 = 255 \; \, MeV$,
 which is lower
than the value $275 \; \, MeV$ found in coherent $\pi^+$ production 
with the $(p, n)$ reaction. Once again the peripheral character of 
the $(p, n)$ reaction should be blamed for it. However, it is 
interesting to note that, in spite of the fact that here we are 
exploring the whole volume of the 
nucleus, like in coherent $\pi^0$ photoproduction, the peak appears 
at higher values of $q^0$ in the neutrino case than in the 
$(\gamma, \pi^0)$ \cite{RCA,LAK} case. In the
latter case the $\Delta$ peak was shifted to energies around 
$190 - 220 \; MeV$
depending on the nucleus. The reason, already discussed in the 
introduction, is the factor $sin^2 \theta$ of coherent $\pi^0$ 
photoproduction which forces finite angles in the cross section 
where the momentum transfer is larger and the nuclear form factor 
smaller.

As we can see, the combination of results of coherent pion production 
induced by photons, neutrinos or hadronic reactions presents 
complementary aspects related to the nuclear properties and the 
propagation of the $\Delta$ and the  pion in the nuclear medium.

In fig. 5 we show the results for the cross section of the same 
reaction integrated now over the electron angles. We observe similar 
features as in fig. 4. The magnitude of the cross section has now 
decreased more with respect to the case where we integrate over the 
pion angles (see figs. 3 and 4), indicating that
the cross section as a function of the electron angle is more forward 
peaked than with the pion angle. This is intuitive since the electron 
momentum is bigger than the pion momentum and a change in angle 
generates larger momentum transfers in the case of the electron, 
which would lead to a larger reduction of the nuclear form factor.

In fig. 6 we show the results for $d \sigma / d E_e$ as a function 
of $q^0$ for different neutrino energies. We can see that the cross 
section increases with the neutrino energy, but at energies above 
$1 GeV$ the increase is more moderate. This is also reminiscent of 
the findings of ref. \cite{FER} in the $(p, n)$ reaction.

In fig. 7 we show the results of $d \sigma / d E_e$ as a function of 
$q^0$, for neutrinos of $1 \; \,  GeV$ scattering from three 
different nuclei, $^{16} O$, $^{37} Cl$ and $^{71} Ga$ used as 
neutrino detectors in several experiments. We observe that the 
cross section grows with A. This is quite different from 
the results found for the 
$(p, n)$ reaction where the cross section decreased from $^{12} C$ to 
$^{40} Ca$ and $^{208} Pb$. The reason for the decrease in the 
hadronic reaction was the distortion of the $p$ and $n$
waves, which does not occur now, since the neutrino and the electron 
are not distorted by the nucleus.
Finally in fig. 8 we show the results for the reaction

\begin{equation}
\nu_{\mu} + A \rightarrow \mu^- + A + \pi^+
\end{equation}

\noindent
corresponding to fig. 7 with electrons.

The evaluation of the cross section in the $\nu_{\mu}$ case is simple 
since both the leptonic tensor and the vector $V^{\mu}$ have the same 
expressions. The only change is in the kinematics in the 
$(\nu_{\mu}, \mu)$ vertex because of the finite mass of the muon. The 
cross sections for the $\nu_{\mu}$ case are decreased by about 
$20 \%$ with respect to those of $\nu_e$ in all nuclei. The reason 
for the decrease is that for a given value of $q^0$ the corresponding 
value of $q$ is larger in the $\mu$ case and hence it leads to larger 
momentum transfers to the nucleus and smaller nuclear form factors. 

Since the Laboratory energetic neutrinos are muon neutrinos, the 
reaction studied here could be implemented with muon neutrinos in 
present Laboratories.

As for the evaluation of cross sections with antineutrinos, the 
changes to be done to obtain them from the ones evaluated in this 
work are minimal, once the transverse parts are neglected as done 
here. One should change

\begin{equation}
\rho_p + \frac{1}{3} \rho_n \rightarrow \rho_n + \frac{1}{3} \rho_p
\end{equation}

\noindent
which in practice amounts to multiplying the neutrino cross 
sections by the factor $( 3 N + Z) / (3 Z + N)$, which is unity for 
isospin symmetric nuclei, as noted in ref. \cite{HKI}.

\section{Conclusions}
We have calculated cross sections for coherent pion production 
in neutrino
(antineutrino) nucleus collisions, of both electron and muon 
type. The calculations have been done accurately taking into 
account the renormalization of the $\Delta$ and pion properties 
in the nuclear medium. We observed that the cross section was 
quite sensitive to these properties, and their inclusion in the 
calculation decreased the cross section by about a factor three 
with respect to the impulse approximation, and shifted the peak 
position to lower excitation energies.

Some of the features, like the shift of the $\Delta$ peak, were 
reminiscent of similar findings in coherent pion production in 
$(^3{He}, t)$ or $(p, n)$ reactions, but the fact that the latter 
are rather peripheral because of the distortion of the hadronic 
beam, by contrast to the neutrino reaction which
occur throughout the nuclear volume, confers the neutrino reaction 
some peculiar features. These features are also different to those 
found in coherent $\pi^0$ photoproduction, also testing the whole 
nuclear volume, because in the latter case there is a factor 
$sin^2 \theta$ in the cross section which reduces the contribution 
of forward angles from where the neutrino cross sections get most 
of their contribution.

All these analogies and differences tell us that the coherent pion 
production induced by neutrinos is an important complement of the 
hadronic and photonuclear processes of pion production in order to 
give information on
pion and $\Delta$ renormalization in a nuclear medium.

On the other hand it is clear that in order to obtain proper 
information about atmospheric neutrinos one has to have a control 
on the different $\nu$-nuclear reactions occuring at intermediate 
energies of the neutrinos, to interpret properly the results of the 
neutrino detectors. The present reaction is one of them.

 In order to test the validity of the model used to obtain the
present results, which can be easily extrapolated to other nuclei 
and other energies, it would be interesting to perform some 
experiment. The cross sections, although small, are in the same 
range as in many experiments performed at present facilities 
\cite{ZEI} and hence are experimentally accessible.

\vspace{1.5cm}

{\bf Acknowledgements:} \\

We would like to acknowledge partial support from CICYT contract no.
AEN 96- 1719. One of us, N. G. K. wishes to acknowledge the 
hospitality of the Valencia University where this work has been 
done.

\newpage
{\large\bf Figure Captions}
\vskip 1cm
\begin{itemize}
\item[1.]
Diagrammatic representation of the coherent pion production
process $\nu_e  + A(g.s) \rightarrow e^- + A(g.s.) + \pi^+$.
\item[2.]
Kinematics of the $\nu n \rightarrow e^- p$ process.
\item[3.]
Angular distribution of the pions from coherent $\pi ^+$ production
on $^{16}O$ with neutrino beam of energy 800 MeV, $T_e=545$ MeV and 
$\theta _e=0^0$. The solid curve corresponds to the full calculation
which includes renormalization of the $\Delta$ and pion in the 
nuclear medium. The dashed curve is the impulse approximation 
calculation.
\item[4.]
Energy spectrum of the coherent pions produced on $^{16}O$ with 
neutrinos of beam energy 1 GeV, and $\theta_e=0^0$. The solid curve 
corresponds to the full calculation with renormalized pions and 
deltas and dashed line is the impulse approximation calculation.
\item[5.]
Same as fig.4, but integrated over the electron solid angle.
\item[6.]
Energy spectra of the coherent pions produced on $^{16}O$, at three
different neutrino energies, with the $\Delta$ and pion 
renormalizations included in the calculations.
\item[7.]
Energy spectra of the coherent pions scattered from three different
nuclei by neutrinos of 1 GeV energy.
\item[8.]
Energy spectra of the coherent pions scattered from three different
nuclei by muon type neutrinos of 1 GeV energy.
\end{itemize}

\end{document}